\documentclass[aps,twocolumn,superscriptaddress,floatfix,longbibliography]{revtex4-2}
\usepackage{amsmath,amssymb,amsthm}
\usepackage{physics}
\usepackage{amsfonts}
\usepackage{mathrsfs}
\usepackage{graphicx}
\usepackage{tabularx}
\usepackage{enumerate}
\usepackage{dcolumn}
\usepackage{bm}
\usepackage{xcolor}
\usepackage[normalem]{ulem}
\usepackage[colorlinks,linkcolor=blue,citecolor=blue,urlcolor=blue]{hyperref}

\begin{document}
\title{Dephasing-induced distinct mobility edges in a dimerized off-diagonal quasicrystal}

\author{Ming-Jie Tao}
\email{taomingjie1020@sina.com}
\affiliation{College of Mathematics and Physics, Chengdu University of Technology, Chengdu 610059, China}

\author{Yi-Ting Wang}
\affiliation{College of Mathematics and Physics, Chengdu University of Technology, Chengdu 610059, China}

\author{Jing Li}
\affiliation{College of Mathematics and Physics, Chengdu University of Technology, Chengdu 610059, China}


\author{Hongsheng Hou}
\affiliation{School of Physics, Hangzhou Normal University, Hangzhou, Zhejiang 311121, China}


\author{Xiang-Ping Jiang}
\email{2015iopjxp@gmail.com}
\affiliation{School of Physics, Hangzhou Normal University, Hangzhou, Zhejiang 311121, China}

\author{Lei Pan}
\email{panlei@nankai.edu.cn}
\affiliation{School of Physics, Nankai University, Tianjin 300071, China}


\date{\today}

\begin{abstract}
Anderson localization and the mobility edge (ME) have been extensively studied in isolated aperiodic systems. Conventional theory suggests that dephasing and decoherence should disrupt localization and facilitate transport. In this work, we investigate localization behaviors in a dimerized off‑diagonal Aubry-Andr\'{e}-Harper (AAH) quasicrystal subject to on‑site pure dephasing. In the strong‑dephasing limit, we apply adiabatic elimination within the Lindblad master‑equation framework to derive an effective classical Markov transition matrix that governs the dissipative relaxation dynamics. Counterintuitively, we demonstrate that pure dephasing can induce distinct MEs, including both conventional MEs separating extended and localized states and anomalous MEs separating multifractal critical states from localized states, even when all eigenstates of the original closed coherent system are delocalized or multifractal. Using fractal dimension finite‑size scaling, wave-packet spreading dynamics, and energy spectrum statistics, we numerically verify the coexistence of fully extended, multifractal critical, and localized regions within the relaxation spectrum of the dissipative system, and construct the global dissipative phase diagram. These findings reveal that dephasing can see as a powerful mechanism for controlling localization transitions, thus enhancing our understanding of dissipative quasicrystal systems. 
\end{abstract}

\maketitle

\section{Introduction}

Anderson localization in disordered quantum systems remains a cornerstone of condensed matter physics~\cite{anderson1958absence}. While arbitrarily weak uncorrelated disorder strictly localizes all single-particle eigenstates in one and two-dimensional (1D and 2D) systems~\cite{thouless1974electrons,abrahams1979scaling,lee1985disordered,kramer1993localization,evers2008anderson}, 1D quasiperiodic lattices, such as the standard Aubry-Andr\'{e}-Harper (AAH) model~\cite{harper1955single,aubry1980analyticity}, can undergo exact localization transitions. Breaking the inherent self-duality of the AAH model typically induces conventional mobility edges (MEs)~\cite{sarma1988mobility,biddle2010pre,ganeshan2015nearest,luschen2018single,liu2018mobility,wang2020one,liu2022anomalous,wang2023exact,li2023observation,qi2023multiple,gonccalves2023critical,hu2025hidden,wang2025family,li2026multifractal,zhou2026fundamental}, critical energy thresholds that segregate extended ergodic states from exponentially localized ones within the single-particle spectrum. Beyond this established paradigm, a more intricate spectral topology emerges in specialized quasiperiodic architectures capable of hosting extended-to-critical or critical-to-localized transitions. In these regimes, the spectrum exhibits anomalous mobility edges (AMEs) that precisely separate multifractal critical states from fully localized ones~\cite{lee2023critical1,zhou2023exact,jiang2024exact1,lu2025exact,wang2025exact}. These anomalous features fundamentally transcend the conventional ME framework, significantly enriching the underlying localization landscape and spurring renewed interest in the exact nature of mobility boundaries in aperiodic systems.

While the aforementioned research extensively maps out localization in isolated, closed quantum systems, realistic experimental platforms are inevitably coupled to their surrounding environments~\cite{luschen2017signatures,xiao2020non,liang2022dynamic,gao2024experimental,yang2026noise,yang2026noise1}. Within the framework of open quantum systems, dissipation and decoherence are traditionally viewed as detrimental factors that disrupt quantum interference. In standard condensed matter systems, on-site pure dephasing is widely expected to act as a delocalized mechanism, suppressing Anderson localization by opening up classical hopping channels and invariably driving the system into a trivial diffusive or sub-diffusive regime~\cite{prosen2008quantum,mebrahtu2012quantum,longhi2019topological,hamazaki2019non,xu2020topological,liuT2020non,shastri2020dissipation,nie2021dissipative,yamamoto2021collective,zeng2020topological1,zeng2020topological2,weidemann2022topological,wu2021non,li2023non,zhu2023topological,liuT2020non,gandhi2023topological,liu2023ergodicity,kawabata2023entanglement,li2024emergent,yu2024non,zhou2024entanglement1,zhou2024entanglement2,jing2024biorthogonal}. Remarkably, recent pioneering theoretical advances have fundamentally challenged this conventional paradigm, revealing that environmental dissipation can counterintuitively stabilize and even actively induce localization transitions~\cite{pan2020non,yamamoto2022universal,mao2023non,mao2024liouvillian,zheng2024exact,ekman2024liouvillian,qin2024occupation,gurvitz2000delocalization,yamilov2014position,huse2015localized,balasubrahmaniyam2020necklace,weidemann2021coexistence,purkayastha2017nonequilibrium,vershinina2017control,yusipov2018quantum,vakulchyk2018signatures,balachandran2019energy,chiaracane2020quasiperiodic,lacerda2021dephasing,chiaracane2022dephasing,dwiputra2021environment,longhi2023anderson,yang2025dissipation,yang2025dissipation1,xu2026dissipation,jiang2026dissipation}. For instance, it has been demonstrated that under strong local dephasing, the rapid relaxation of quantum coherence allows for the adiabatic elimination of off-diagonal density matrix elements, transforming the quantum dynamics into a classical Markov master equation~\cite{yusipov2017localization,liu2024dissipation,longhi2024dephasing}. Under specific conditions, the spectral indicators of this effective relaxation matrix can manifest extended-to-localized transitions and dissipative MEs, unlocking a completely new avenue for controlling quantum localization and transport via dephasing engineering.

In this work, we investigate localization properties in a 1D dimerized off-diagonal AAH quasicrystal under dephasing, in which we demonstrate the coexistence MEs and AMEs in the spectrum as a function of the onsite potential strength. By employing the Lindblad master equation to derive the effective Markov transition matrix in the strong-dephasing regime, we reveal our central finding: pure dephasing can dynamically induce a remarkably rich localization transitions within a single system. Specifically, we demonstrate that dephasing can induce the localization transitions and generates distinct MEs. This mechanism cleanly segregates fully extended, critically multifractal, and localized states within the identical relaxation spectrum. Furthermore, the resulting global dissipative phase diagram starkly contrasts with its purely coherent counterpart, establishing pure dephasing not merely as a source of decoherence, but as an active, powerful tool for reshaping localization boundaries and manipulating quantum transport.

The remainder of this paper is organized as follows. In Sec.~\ref{sec:2}, we introduce the tight-binding Hamiltonian of the dimerized off-diagonal quasicrystal and define the diagnostic indicators, including the fractal dimension (FD) and dynamical variance. In Sec.~\ref{sec:3}, we formulate the Lindblad master equation and detail the adiabatic elimination leading to the effective classical Markov matrix. Sec.~\ref{sec:4} presents our numerical results, detailing the static fractal scaling, transport dynamics, and level statistics that verify the MEs and AMEs. Finally, we summarize our main conclusions and offer an experimental outlook in Sec.~\ref{sec:5}.

\section{The model Hamiltonian}\label{sec:2}

We consider a 1D tight-binding model of free spinless fermions subjected to both lattice dimerization and an off-diagonal quasiperiodic modulation. The system under consideration is governed by the Hamiltonian
\begin{equation}\label{equation1}
    H = \sum_{n=1}^{L} t_n \left( \hat{c}_{n}^{\dagger} \hat{c}_{n+1} + \hat{c}_{n+1}^{\dagger} \hat{c}_{n} \right),
\end{equation}
where $\hat{c}_{n}^{\dagger}$ ($\hat{c}_{n}$) is the fermionic creation (annihilation) operator at the $n$-th lattice site, satisfying standard anti-commutation relations, and $L$ denotes the total number of lattice sites. The spatially varying nearest-neighbor hopping amplitude, $t_n$, encapsulates both a deterministic dimerization and an incommensurate quasiperiodic order, formulated as
\begin{equation}\label{equation2}
    t_n = t + \lambda \cos(\pi n) + V \cos(2\pi\alpha n + \phi).
\end{equation}

\begin{figure}[t!]
	\centering
	\includegraphics[width=0.48\textwidth]{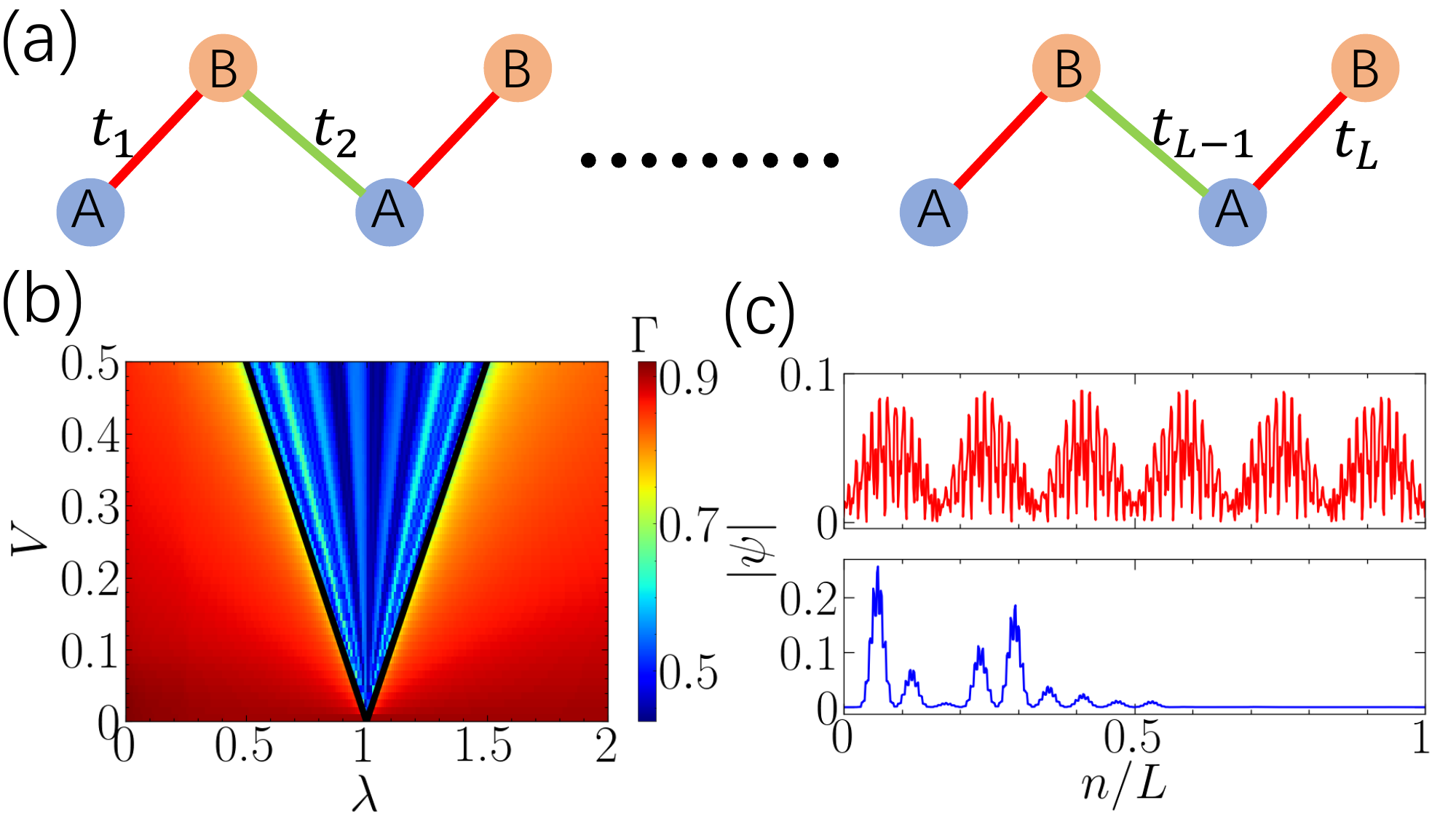}
        \includegraphics[width=0.48\textwidth]{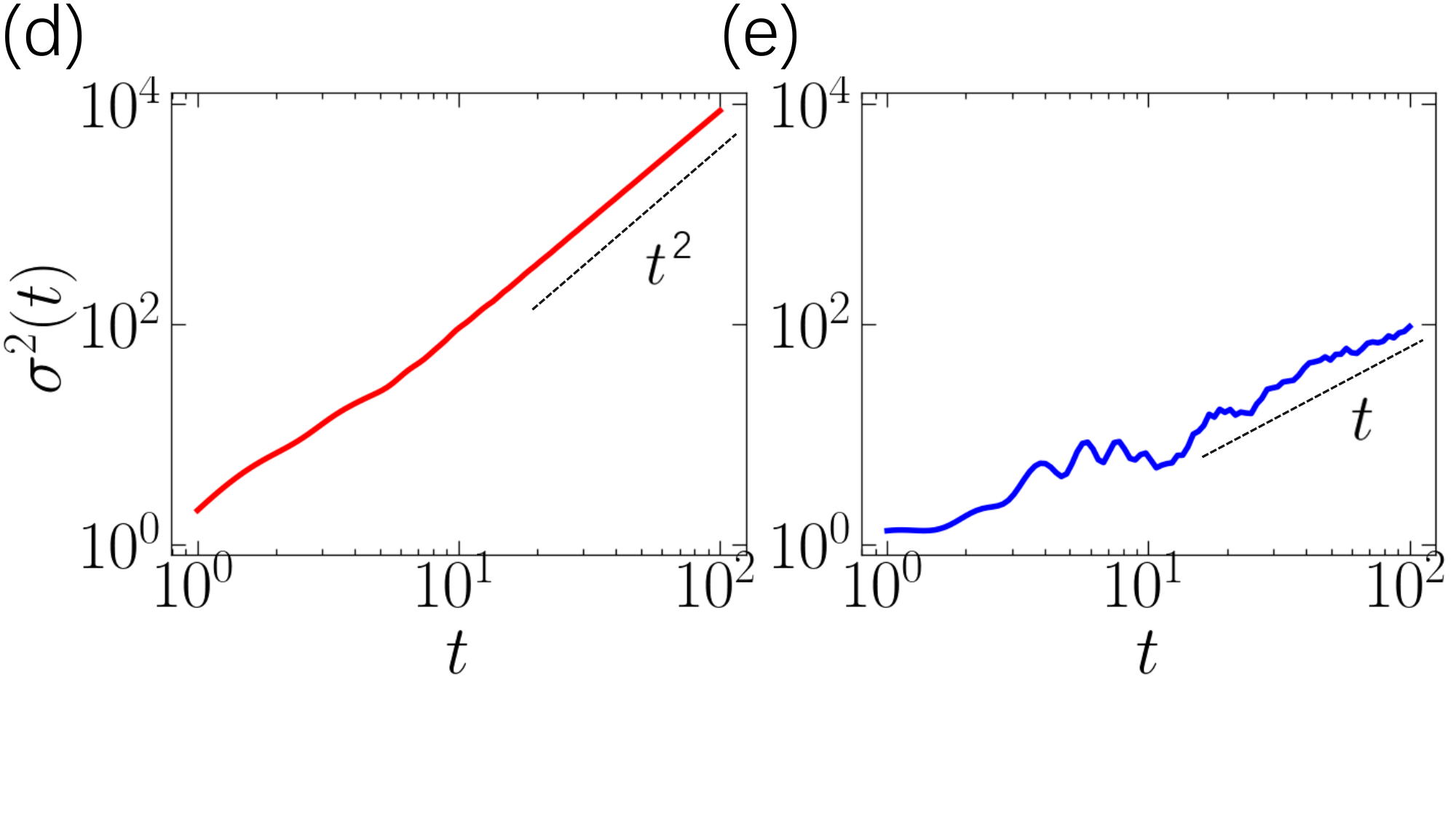}
        \vspace{-0.5in}
        \caption{(a) Schematic illustration of the 1D Hamiltonian governed by Eq.~(\ref{equation1}). The red and green solid lines denote the intra- and inter-cell hopping amplitudes between sublattices $A$ and $B$, respectively. (b) Global phase diagram in the $(\lambda, V)$ parameter space characterized by the fractal dimension $\Gamma$ (in units of $t=1$). The red and blue regions designate the fully extended and multifractal critical phases, respectively. The solid black curves represent the exact boundaries separating these two distinct regimes. (c) Real-space probability distributions of representative single-particle eigenstates, visually contrasting the extended and critical states. (d)-(e) Time evolution of the wave-packet variance $\sigma^2(t)$, exhibiting rapid ballistic expansion ($\sigma^2(t) \propto t^2$) in the extended phase and diffusive spreading ($\sigma^2(t) \propto t$) in the critical phase.}
	\label{fig1}
\end{figure}

In this framework, the parameters are meticulously defined to govern the competition between different spectral phases, see Fig.~\ref{fig1}(a). The parameter $t$ represents the unperturbed uniform hopping amplitude, which we set to $t=1$ to establish the fundamental energy scale for all subsequent analyses. The term $\lambda \cos(\pi n)$ introduces a dimerization, yielding alternating intra-cell and inter-cell hopping strengths of $t-\lambda$ and $t+\lambda$, respectively. The quasiperiodic off-diagonal modulation is governed by the amplitude $V$ and an irrational frequency $\alpha$. To rigorously implement periodic boundary conditions in our numerical simulations, the lattice size is chosen as a Fibonacci number, $L=F_m$, defined recursively by $F_{m+1}=F_{m}+F_{m-1}$ with $F_0=F_1=1$. Consequently, the irrational frequency is given by the Diophantine approximation of the golden mean, $\alpha = \lim_{m\to\infty} F_{m-1}/F_m = (\sqrt{5}-1)/2$. The phase offset $\phi$ can be set arbitrarily or averaged over to remove specific boundary artifacts.

To quantitatively characterize the localization properties of the single-particle eigenstates $|\psi^{(k)}\rangle = \sum_{n} \psi_n^{(k)} \hat{c}_n^\dagger |0\rangle$, we employ the inverse participation ratio (IPR) and the corresponding fractal dimension. For a normalized eigenstate, the IPR is defined as
\begin{equation}\label{equation3}
    \mathrm{IPR}_k = \sum_{n=1}^{L} \left| \psi_n^{(k)} \right|^4.
\end{equation}
The scaling behavior of the IPR with the system size $L$ is captured by the fractal dimension $\Gamma_k$, defined as
\begin{equation}\label{equation4}
    \Gamma_k = -\lim_{L\to\infty} \frac{\ln (\mathrm{IPR}_k)}{\ln L}.
\end{equation}
The index $\Gamma_k$ serves as a robust indicator for the localization nature of the wave function: in the thermodynamic limit $\Gamma_k \to 0$ signifies an exponentially localized state, $\Gamma_k \to 1$ rigorously identifies an extended (ergodic) state, and fractional values $0 < \Gamma_k < 1$ are the hallmark of a critical (multifractal) state. Beyond static spectral properties, the dynamical transport signatures of the system are extracted by observing the temporal spreading of an initially localized single-site excitation. In Figs.~\ref{fig1}(b) and (c), we show the phase diagram of the Hamiltonian (\ref{equation1}) and representative extended and multifractal critical states, respectively.. The wave-packet expansion is characterized by the time evolution of the second moment (variance) of the spatial distribution,
\begin{equation}\label{equation5}
    \sigma^2(t) = \sum_{n=1}^{L} (n-n_0)^2 |\psi_n(t)|^2,
\end{equation}
where $\psi_n(t)$ is the probability amplitude at site $n$ and time $t$, and $n_0$ is the initial excitation site. In the asymptotic long-time limit, the variance exhibits a power-law growth, $\sigma^2(t) \propto t^\delta$. The dynamical exponent $\delta$ provides a macroscopic diagnostic of the underlying phase:
\begin{equation}\label{equation6}
    \delta = 
    \begin{cases} 
        0, & \text{localized regime,} \\ 
        1, & \text{critical (diffusive) regime,} \\ 
        2, & \text{extended (ballistic) regime.} 
    \end{cases}
\end{equation}
In Figs.~\ref{fig1}(d) and (e), we show the dynamical evolution of representative extended and critical states, respectively, in agreement with Eq.~(\ref{equation6}). By systematically varying the dimerization $\lambda$ and quasiperiodic strength $V$ (e.g., $\lambda=0.2$ and $1.0$ in our typical cases), we map out the coherent phase diagram. This establishes the baseline from which we subsequently investigate the localization transitions induced by strong environmental dephasing ($\gamma \gg t$).

\section{Dephasing-induced distinct mobility edges}\label{sec:3}

To investigate the impact of environmental noise on the localization properties of our system, we model the dephasing effects utilizing the open quantum system formalism. The Lindblad master equation for the density matrix $\hat{\rho}(t)$ describing dephasing effects reads~\cite{lindblad1976generators}:
\begin{equation}
    \frac{d\hat{\rho}(t)}{dt} \equiv \mathcal{L}[\hat{\rho}(t)] = -i[\hat{H}, \hat{\rho}(t)] + \mathcal{D}[\hat{\rho}(t)],
    \label{eq:Lindblad}
\end{equation}
where $\mathcal{L}$ denotes the Liouvillian superoperator, and $\mathcal{D}[\hat{\rho}(t)]$ represents the dissipator that captures the non-unitary environmental coupling. The dissipator is explicitly formulated as
\begin{equation}
    \mathcal{D}[\hat{\rho}(t)] = \sum_{n} \gamma_{n} \left( \hat{l}_{n} \hat{\rho} \hat{l}_{n}^{\dagger} - \frac{1}{2} \{ \hat{l}_{n}^{\dagger} \hat{l}_{n}, \hat{\rho} \} \right),
    \label{eq:dissipator}
\end{equation}
where $\hat{l}_{n}$ represents the quantum jump operators associated with the local dissipation channel at site $n$, and $\gamma_{n}$ denotes the corresponding dephasing strength. Here, we restrict our attention to pure, spatially homogeneous on-site dephasing by choosing $\hat{l}_{n} = \hat{c}_{n}^{\dagger}\hat{c}_{n}$ and setting $\gamma_{n} \equiv \gamma > 0$. Because the jump operators commute with the total particle number operator, this local dephasing mechanism strictly conserves the total number of excitations in the system. 

In the single-particle sector of the Hilbert space spanned by the set of states $|n\rangle = \hat{c}_{n}^{\dagger}|0\rangle$, the matrix elements of the density matrix are defined as $\rho_{n,m}(t) = \langle n|\hat{\rho}(t)|m\rangle$. By evaluating Eq.~\eqref{eq:Lindblad}, the coupled equations of motion for the populations ($n=m$) and the quantum coherence ($n \neq m$) yield
\begin{align}
    \frac{d\rho_{n,m}}{dt} = & \, i \left( t_{n-1}\rho_{n-1,m} + t_{n}\rho_{n+1,m} \right) \nonumber \\
    & - i \left( t_{m}\rho_{n,m+1} + t_{m-1}\rho_{n,m-1} \right) \nonumber \\
    & - \gamma (1 - \delta_{n,m}) \rho_{n,m},
    \label{eq:matrix_elements}
\end{align}
where $t_n$ represents the quasiperiodic hopping amplitudes defined in Eq.~\eqref{equation2}.

\begin{figure*}[t]
	\centering
	\includegraphics[width=0.320\textwidth]{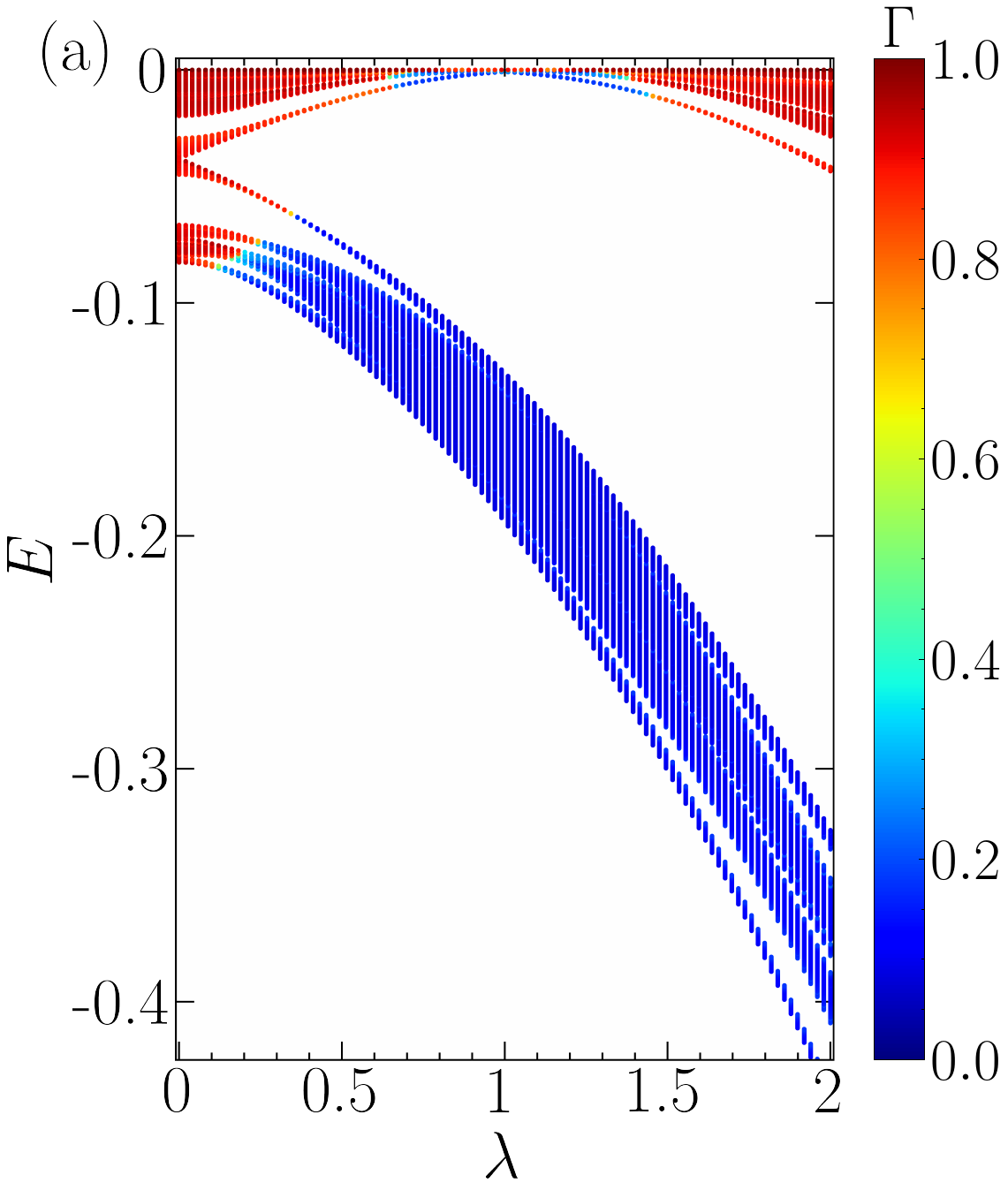}
	\includegraphics[width=0.325\textwidth]{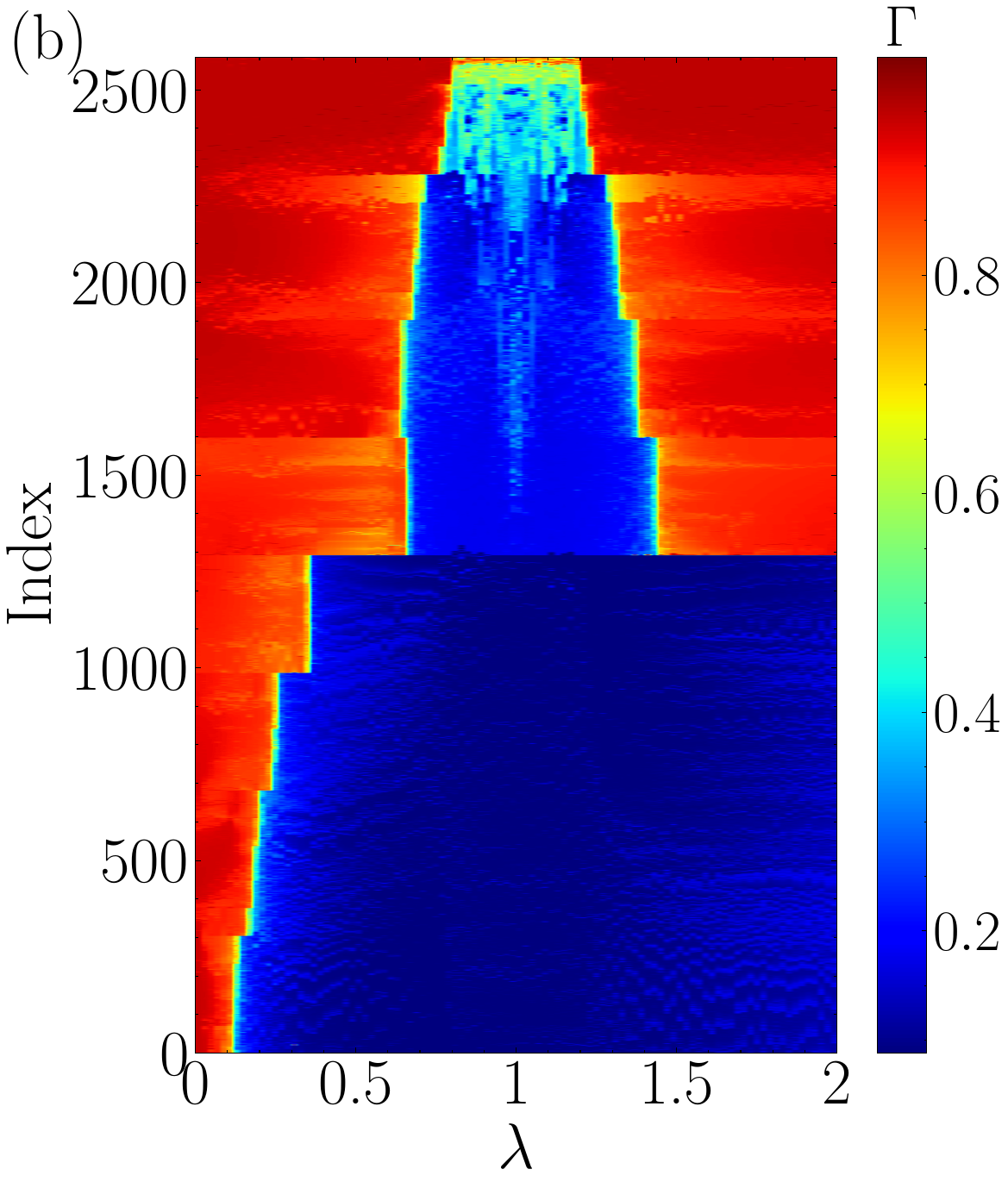}
        \includegraphics[width=0.315\textwidth]{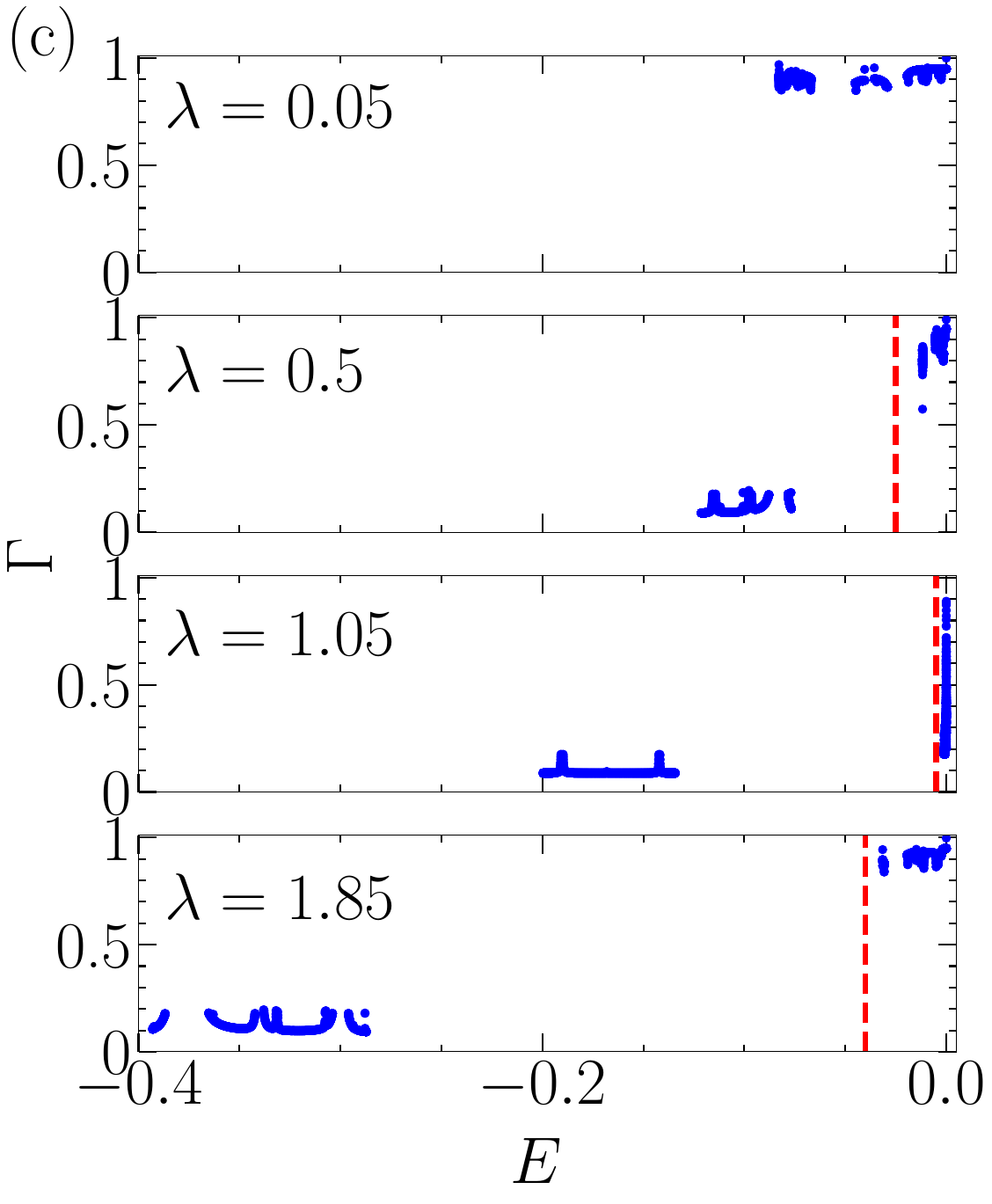}
        \caption{Spectral properties and localization transitions of the effective Markov transition matrix $M$ under dephasing. (a) The eigenvalue $E$ as a function of the dimerization strength $\lambda$. The color gradient encodes the fractal dimension $\Gamma$ of the corresponding single-particle eigenstates, enabling a clear visual distinction between the fully extended ($\Gamma \to 1$, red), critically multifractal, and exponentially localized ($\Gamma \to 0$, blue) states. (b) The fractal dimension $\Gamma$ for all macroscopic eigenstates (sorted ascendingly by their eigenvalue index) versus $\lambda$. This global map illustrates the macroscopic population fractions of the distinct regions and the structural evolution of the phase boundaries. (c) The fractal dimension $\Gamma$ projected onto the eigenvalue $E$ for four representative modulation strengths: $\lambda = 0.05, 0.5, 1.05,$ and $1.85$. The vertical red dashed lines precisely denote the locations of the conventional mobility edges (MEs) and anomalous mobility edges (AMEs), which act as strict thresholds segregating the extended, critical, and localized states within the identical relaxation spectrum. The fixed system parameters for all panels are set to $V = 0.2$, the uniform dephasing rate $\gamma = 100$, and the lattice size $L = 2584$.}
	\label{fig2}
\end{figure*}

The central objective of this work is to determine whether environmental dephasing can counterintuitively induce localization transitions and give rise to distinct MEs. To address this problem analytically, we focus on the strong-dephasing  regime where $\gamma \gg \{t, \lambda, V\}$. In this limit, a significant separation of timescales occurs: the off-diagonal coherence terms $\rho_{n,m}$ ($n \neq m$) decay rapidly on a short characteristic timescale of $\sim \gamma^{-1}$, whereas the diagonal populations $P_{n}(t) \equiv \rho_{n,n}(t)$ evolve on a much slower macroscopic timescale. This allows for the systematic adiabatic elimination of the fast coherence variables by setting $d\rho_{n,m}/dt \approx 0$ for $n \neq m$. Substituting the steady-state coherences back into the population equations reduces the full quantum dynamics to a classical master equation governing the probability distribution:
\begin{equation}
    \frac{dP_{n}}{dt} = \sum_{m=1}^{L} M_{n,m} P_{m},
    \label{eq:classical_master}
\end{equation}
where the elements of the Markov transition matrix $M$ are derived as
\begin{equation}
    M_{n,m} = \frac{2}{\gamma} \left[ t_{n}^{2}\delta_{n,m-1} + t_{n-1}^{2}\delta_{n,m+1} - (t_{n}^{2} + t_{n-1}^{2})\delta_{n,m} \right].
    \label{eq:Markov_matrix}
\end{equation}
This effective classical master equation can also be physically unraveled as a single-particle Schrödinger equation subjected to stochastically frequent, periodic phase randomizations occurring at uniform time intervals of $\Delta t = 2/\gamma$. Crucially, because $M$ is a real symmetric (Hermitian) matrix, Eq.~\eqref{eq:classical_master} can be mathematically mapped onto a standard Schrödinger equation with a Wick-rotated, anti-Hermitian Hamiltonian $\hat{H}' = iM$. Let $E_k$ and $\Psi_n^{(k)}$ ($k=1,2,\dots,L$) denote the eigenvalues and corresponding orthonormal eigenvectors of $M$. Due to the conservation of total probability, the eigenvalues are strictly non-positive and can be ordered as $E_1 \le E_2 \le \dots \le E_L = 0$. While the decaying eigenstates associated with non-vanishing eigenvalues ($E_k < 0$) do not describe physical steady states individually, they provide a mathematically complete basis for the full relaxation dynamics. The system always possesses a unique invariant steady state at $E_L = 0$, characterized by the uniformly extended, maximally mixed eigenvector $\Psi_n^{(L)} = (1/L)(1,1,\dots)^T$. Consequently, any initial localized wave packet will ultimately diffuse and homogenize across the entire lattice in the long-time limit ($t \to \infty$).

\begin{figure}[t]
	\centering
	\includegraphics[width=0.48\textwidth]{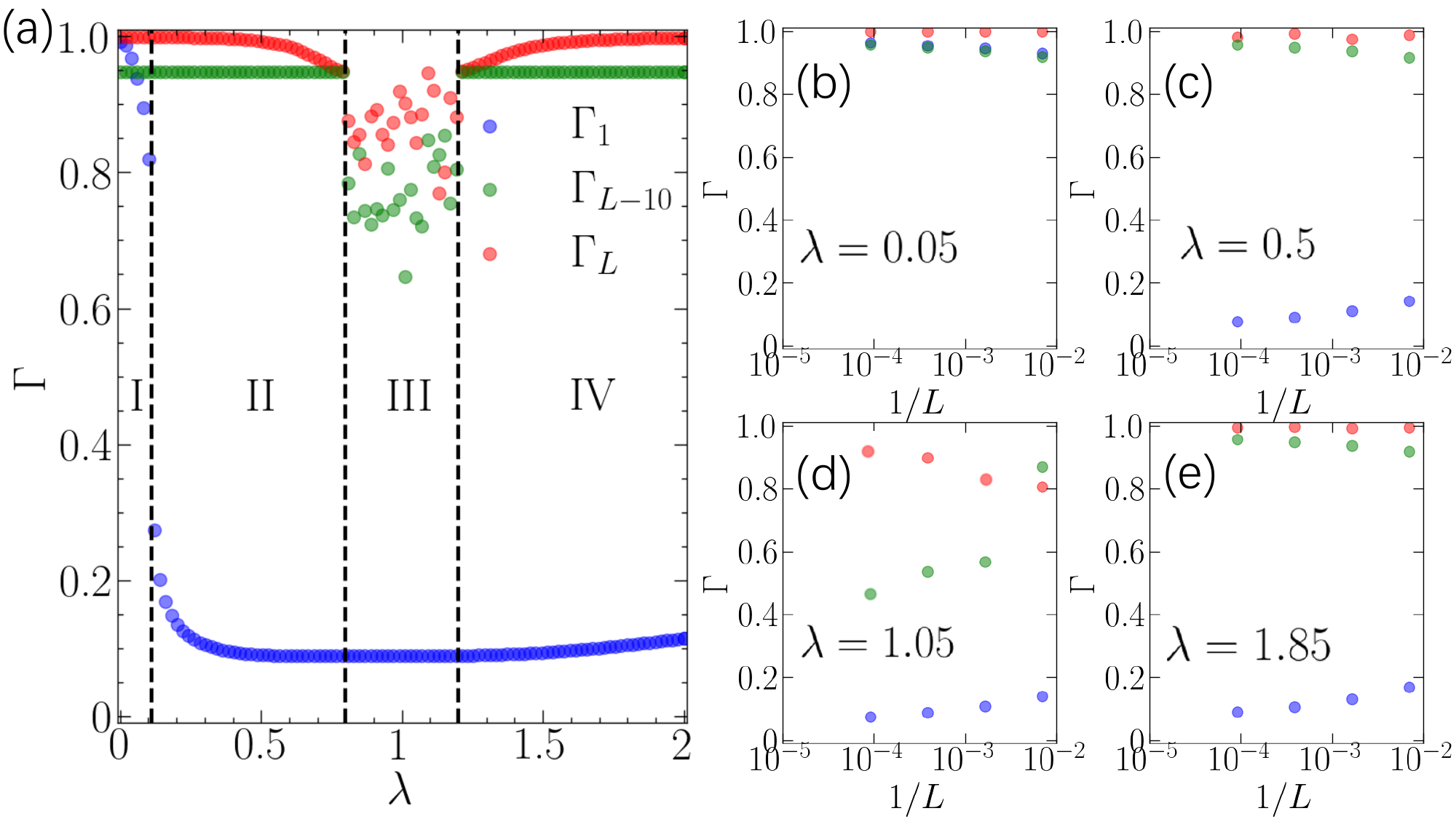}
        \caption{Finite-size scaling analysis of the fractal dimension $\Gamma$ for representative single-particle eigenstates. The scaling behaviors of $\Gamma$ as a function of $1/L$ are presented under four distinct modulation strengths: (a) $\lambda = 0.05$, (b) $\lambda = 0.5$, (c) $\lambda = 1.05$, and (d) $\lambda = 1.85$. By extrapolating to the thermodynamic limit ($L \to \infty$, corresponding to $1/L \to 0$), the distinct nature of the eigenstates is rigorously verified. Specifically, the fully extended states (red squares) robustly converge to $\Gamma \to 1$, the exponentially localized states (blue triangles) decay to $\Gamma \to 0$, and the critically multifractal states (green diamonds) converge to intermediate fractional values ($0 < \Gamma < 1$). The fixed system parameters are $V = 0.2$ and $\gamma = 100$.}
	\label{fig3}
\end{figure}

\begin{figure}[t]
	\centering
	\includegraphics[width=0.48\textwidth]{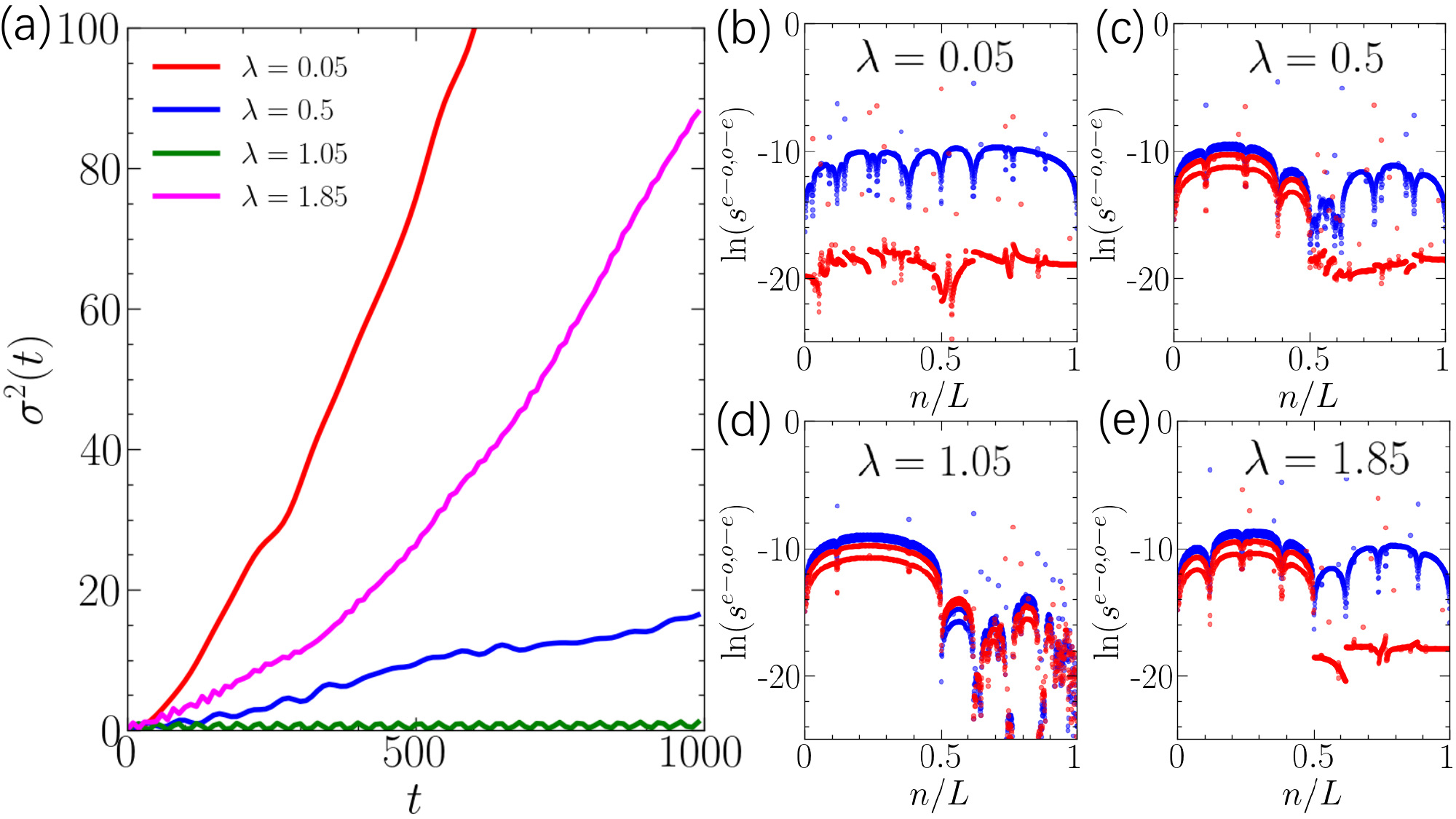}
    \caption{Dynamical wave-packet transport and spectral statistics characterizing the distinct regions. (a) Time evolution of the spatial variance $\sigma^2(t)$ under selected dimerization strength $\lambda$. The distinct temporal growth profiles directly manifest the underlying quantum transport mechanisms: robust ballistic/superdiffusion expansion in the extended phase and ME region, intermediate subdiffusive spreading in the AME region, and a complete transport arrest ($\sigma^2(t) \sim \mathrm{const}$) verifying the exponentially localized phase. (b)-(e) Logarithmic distributions of the adjacent even-odd, $\ln(s^{e-o})$, and odd-even, $\ln(s^{o-e})$, level spacings as a function of the normalized state index $n/L$. These spectral spacing profiles are evaluated at $\lambda = 0.05, 0.5, 1.05$, and $1.85$, respectively. The contrasting statistical signatures of the energy gaps serve as an independent, highly sensitive diagnostic tool to unambiguously corroborate the phase boundaries of the extended, critical, and localized regimes.}
	\label{fig4}
\end{figure}

Nevertheless, the transient relaxation and transport dynamics can be dramatically hindered if the Markov matrix $M$ hosts localized eigenvectors whose corresponding eigenvalues lie extremely close to zero, thereby possessing anomalously long lifetimes. At this juncture, we highlight a fundamental distinction between diagonal and off-diagonal quasiperiodic systems under dephasing noise. In conventional models where the quasiperiodic potential is purely diagonal (on-site) and the hoppings are uniform ($t_n \equiv t$), the transition rates in Eq.~\eqref{eq:Markov_matrix} become spatially homogeneous. In that scenario, as demonstrated in Ref.~\cite{ref24}, all eigenvectors of $M$ are strictly extended, meaning dephasing invariably destroys localization and enhances diffusion. In stark contrast, for our dimerized off-diagonal quasicrystal model, the incommensurate modulation is embedded directly within the hopping amplitudes $t_n$. As a consequence, the classical transition rates themselves inherit the quasiperiodic structure. This competition between diagonal and off-diagonal quasiperiodicity in the effective classical matrix $M$ opens up the possibility for non-trivial localization transitions and the formation of distinct MEs within the relaxation spectrum.

\section{Numerical results and discussions}\label{sec:4}

According to Eq.~\eqref{eq:Markov_matrix}, the effective classical Markov matrix $M$ simultaneously incorporates both diagonal and off-diagonal quasiperiodic modulations. The intricate competition between these two terms is anticipated to precipitate non-trivial localization transitions and the emergence of distinct MEs. To substantiate this physical picture, we numerically diagonalize the matrix $M$ under periodic boundary conditions (PBC) and systematically analyze both the static properties and the dynamical behaviors of the corresponding eigenstates.

To quantitatively map the localization transitions, and the three different typical extended, critical and localized states in the quasiperiodic system, we compute the fractal dimension $\Gamma$ of the eigenstates across the energy spectrum as the function of the dimerization strength $\lambda$ with the fixed $V=0.2$, as presented in Fig.~\ref{fig2}(a). The numerical results unambiguously demonstrate that MEs are dynamically generated when the quasiperiodic modulation surpasses a critical threshold. Furthermore, we show in Fig.~\ref{fig2}(b) the $\Gamma$ for all macroscopic eigenstates
(sorted ascendingly by their eigenvalue index) versus $\lambda$. This global map illustrates the macroscopic population fractions of the distinct regions and the structural evolution of the phase boundaries. For fixed $V=0.2$, one can clearly observe that the system lies entirely in the extended-state regime for $\lambda<0.12$. A ME separating extended and localized states emerges for $0.12<\lambda<0.8$. In the interval $0.8<\lambda<1.2$, an AME appears that distinguishes critical multifractal states from localized states. Upon further increasing $\lambda$, the system re-enters the region where a ME is present. As shown in Fig.~\ref{fig2}(c), we show four representative parameter regimes of the dissipative system for $\lambda=0.05$, $0.5$, $1.05$, and $1.85$, corresponding to the fully-extended regime, the ME‑hosting regime, the AME‑hosting regime, and again the ME‑hosting regime, respectively. These numerical results indicate that pure dephasing can induce localization transitions for the Hamiltonian (\ref{equation1}), despite the original system being in extended or critical phases. In Fig.~\ref{fig3}(a), we select three representative eigenstates and plot their fractal dimensions $\Gamma_1$, $\Gamma_{L-10}$, and $\Gamma_L$, sorted by ascending energy. From these results, we can identify fully extended, multifractal critical, and exponentially localized states. Furthermore, the finite-size scaling analysis of $\Gamma$ [Figs.~\ref{fig2}(b)-(e)] confirms the rigorous coexistence of exponentially localized ($\Gamma \to 0$), critically multifractal ($0 < \Gamma < 1$), and fully extended ergodic ($\Gamma \to 1$) states within the identical spectrum. This spectral segmentation is strictly mediated by AMEs, denoted by the dashed lines, which theoretically separate these  distinct regimes.

To further probe the transport signatures distinguishing these novel phases, we simulate the non-equilibrium spreading dynamics of an initially localized single-site wave packet. Figure~\ref{fig4}(a) illustrates the time evolution of the spatial variance $\sigma^2(t)$. In the extended phase, the wave packet exhibits rapid ballistic expansion ($\sigma^2(t) \propto t^2$), whereas in the critical regime, the transport is significantly hindered, manifesting as sub-diffusive or diffusive spreading ($\sigma^2(t) \propto t$). Conversely, in the strongly localized regime, $\sigma^2(t)$ rapidly saturates to a constant finite value, signaling a complete arrest of transport induced by pure dephasing. 

Additionally, the phase boundaries can be statistically determined by analyzing adjacent energy-level spacings, specifically the even‑odd ($\delta_n^{e-o} = E_{2n}-E_{2n-1}$) and odd‑even ($\delta_n^{o-e} = E_{2n+1}-E_{2n}$) gap statistics. In the extended phase, near‑double degeneracy in the spectrum causes $\delta_n^{e-o}$ to vanish, giving rise to a clear, experimentally observable gap between the two spacing distributions, as shown in Fig.~\ref{fig4}(b). In the localized phase, this gap closes completely, with $\delta_n^{e-o}$ and $\delta_n^{o-e}$ becoming nearly identical. Notably, the critical phase exhibits a characteristic scattered, scale‑invariant distribution, yielding a robust statistical signature fundamentally distinct from those of both the extended and localized phases~\cite{deng2019one,zhang2022localization,sarkar2021mobility}. For Figs.~\ref{fig4}(c) and (e), the level spacing statistics exhibit features of both extended and localized states, indicating that these parameter points lie in the regime hosting a ME. By contrast, the level spacing statistics in Fig.~\ref{fig4}(d) show signatures of critical and localized states, corresponding to the parameter regime with an AME.

\begin{figure}[!t]
	\centering
	\includegraphics[width=0.48\textwidth]{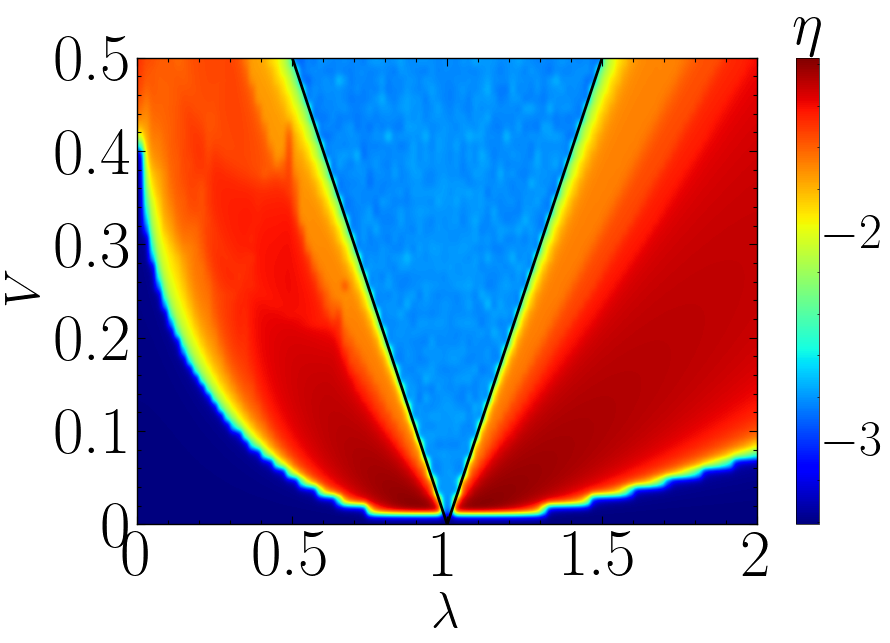}
    \caption{Global phase diagram of the dissipative system in the $(\lambda,V)$ parameter space in the strong‑dephasing regime. Boundaries mark three distinct spectral regimes: the fully‑extended regime, a mixed regime of extended and localized states, and a mixed regime of critical and localized states, separated by conventional MEs and AMEs. This dissipative phase diagram contrasts sharply with the purely coherent phase diagram in Fig.~\ref{fig1}(a), giving unambiguous evidence that environmental dephasing counterintuitively reshapes the localization landscape.}
	\label{fig5}
\end{figure}

To characterize the dissipative phase diagram of the system under dephasing, we introduce the averaged inverse participation ratio (IPR) and the normalized participation ratio (NPR). The NPR is defined as ${\rm NPR}_k=\bigl(L\sum_{n} |\psi_n^{(k)}|^{4}\bigr)^{-1}$. Averaging over all eigenstate values $\{{\rm IPR}_k\}$ and $\{{\rm NPR}_k\}$, we obtain
\begin{equation}\label{equation12}
\langle {\rm IPR} \rangle=\frac{1}{L}\sum_{k=1}^L {\rm IPR}_k, \quad
\langle {\rm NPR} \rangle=\frac{1}{L}\sum_{k=1}^L {\rm NPR}_k.
\end{equation}
In the thermodynamic limit $L\to\infty$, these averaged quantities provide criteria for different phases:
the system resides in the extended phase when $\langle{\rm IPR}\rangle \simeq 0$ and $\langle{\rm NPR}\rangle$ remains finite;
it is in the localized phase when $\langle{\rm IPR}\rangle$ is finite and $\langle{\rm NPR}\rangle \simeq 0$;
and it occupies the intermediate multifractal critical phase when both $\langle {\rm IPR} \rangle$ and $\langle {\rm NPR} \rangle$ take finite values.
Building upon $\langle{\rm IPR}\rangle$ and $\langle{\rm NPR}\rangle$, we compute the composite diagnostic quantity $\eta$~\cite{roy2021reentrant,qi2023multiple},
\begin{equation}\label{equation13}
\eta=\log_{10}\bigl[\langle {\rm IPR} \rangle \times \langle {\rm NPR} \rangle\bigr].
\end{equation}
This quantity $\eta$ enables us to clearly separate the intermediate region from fully extended and fully localized domains within the phase diagram, and further distinguishes parameter regimes governed by conventional MEs and AMEs. The resulting global dissipative phase diagram is presented in Fig.~\ref{fig5}. Plotted in the $(\lambda,V)$ parameter plane, this map clearly delineates well‑defined boundaries separating the fully‑extended regime, the ME‑hosting regime coexisting extended and localized states, and the AME‑hosting regime containing critical and localized states. Taken together, this phase diagram yields conclusive numerical evidence that, contrary to conventional expectations, pure dephasing within this dimerized off‑diagonal quasicrystal can actively drive localization transitions and stabilize distinct ME phases across a broad parameter window.

\section{Conclusion}\label{sec:5}

In summary, we have investigated localization properties in a 1D dimerized off-diagonal AAH quasicrystal subject to homogeneous on-site pure dephasing. In the strong-dephasing regime, we maps the Lindblad dynamics onto an effective classical Markov process whose spatially dependent transition rates inherit the quasiperiodic structure of the hopping amplitudes. Remarkably, the resulting relaxation spectrum supports fully extended, multifractal critical, and exponentially localized modes within the same spectrum, even when the corresponding closed coherent system contains only extended or critical states. Finite-size scaling of the fractal dimension, wave-packet dynamics, and energy-level statistics consistently confirm the coexistence of these distinct regimes and their separation by conventional MEs and AMEs. The emergence of these distinct MEs originates from the interplay between dephasing and off-diagonal quasiperiodicity: dephasing converts the quasiperiodically modulated hopping into spatially inhomogeneous classical transition rates, thereby reshaping the localization properties of the relaxation modes rather than simply suppressing localization. The resulting dissipative phase diagram exhibits extended, ME, and AME regimes and is qualitatively distinct from that of the corresponding closed system. Our results demonstrate that pure dephasing can serve as an effective mechanism for engineering localization transitions and distinct MEs in quasiperiodic open systems.

\section*{Acknowledgments}

This work is supported by the National Natural Science Foundation of China (Grants No.~12304388, No.~12304290, and No.~12505017), the Beijing National Laboratory for Condensed Matter Physics (Grant No.~2025BNLCMPKF017), and the Fundamental Research Funds for the Central Universities.

\appendix

\renewcommand{\thesection}{\Alph{section}}
\renewcommand{\thefigure}{A\arabic{figure}}
\renewcommand{\thetable}{A\Roman{table}}
\setcounter{figure}{0}
\renewcommand{\theequation}{A\arabic{equation}}
\setcounter{equation}{0}

\bibliography{Localization}
\end{document}